\documentclass[final,1p,times]{elsarticle}

\usepackage{amssymb}
\usepackage{lipsum}
\usepackage[rgb]{xcolor}
\usepackage[version=4]{mhchem}

\biboptions{sort&compress}

\usepackage[draft]{changes} 
\definechangesauthor[color=blue]{NB}

\usepackage{hyperref}
\hypersetup{
    colorlinks,%
    citecolor=blue,%
    filecolor=black,%
    linkcolor=magenta,%
    urlcolor=black
}

\journal{Acta Materialia}

\begin{document}

\begin{frontmatter}



\title{Ductility and Brittle Fracture of Tungsten by Disconnection Pile-up on Twin Boundaries}
 
\author[inst1,inst2]{Omar Hussein\corref{cor}}
\cortext[cor]{Corresponding author}
\ead{ohussein@lanl.gov}
\author[inst1]{Nicolas Bertin}
\author [inst3]{Jakub Veverka}
\author[inst1]{Tomas Oppelstrup}
\author [inst4]{Jaime Marian}
\author[inst5,inst6]{Fadi Abdeljawad}
\author[inst3]{Shen J. Dillon}
\author[inst1]{Timofey Frolov\corref{cor}}
\cortext[cor]{Corresponding author}
\ead{frolov2@llnl.gov}

\affiliation[inst1]{Lawrence Livermore National Laboratory, Livermore, CA 94550, USA}
\affiliation[inst2]{Los Alamos National Laboratory, Los Alamos, NM 87545, USA}
\affiliation[inst3]{Department of Materials Science and Engineering, University of California Irvine, Irvine, CA, 92697}
\affiliation[inst4]{Department of Materials Science and Engineering, University of California Los Angeles, Los Angeles, CA, 90095}
\affiliation[inst5]{Department of Materials Science and Engineering, Lehigh University, Bethlehem, PA 18015, USA}
\affiliation[inst6]{Department of Materials Science and Engineering, Northwestern University, Evanston, IL 60208, USA}

\begin{abstract}
Refractory body-centered cubic (BCC) metals and alloys are of extraordinary importance in modern technological and structural applications. However, their wider adoption in science and technology is severely restricted by low-temperature brittleness, quantified by an unacceptably high value of the brittle-to-ductile transition temperature (DBTT). The DBTT of these alloys is known to depend strongly on the particular microstructure of the material following mechanisms that are not well understood. Here we apply cross-scale molecular dynamics (MD), a simulation approach that preserves full atomic resolution while capturing the collective evolution of dislocations, twins, and cracks in near-micron-scale volumes, to investigate ductility and fracture in single-crystal tungsten pillars as a function of initial defect microstructure, deformation conditions, and temperature. The simulations reveal a sequence of microscopic processes conducive to failure: dislocation starvation, nucleation and growth of twins, pinning of the twin boundaries at surface asperities, resulting in disconnection pile-ups that trigger crack nucleation and propagation at low macroscopic stresses along incoherent boundary segments.
By resolving these processes within a single atomistic framework, our simulations connect defect-level dynamics to macroscopic fracture behavior and identify microstructural pathways capable of shifting the DBTT through targeted promotion or suppression of the underlying deformation mechanisms.
\end{abstract}





\end{frontmatter}

\section{Introduction}

Refractory metals and their alloys exhibit an exceptional combination of physical and mechanical properties, including extraordinary melting point, high elastic moduli, and superior resistance to creep and wear degradation at extreme temperatures~\cite{SNEAD2019585}. These attributes render them indispensable in many demanding applications, from aerospace propulsion and gas turbines to rocket nozzles, nuclear energy systems, high-current electrical contacts, heating elements, and high-temperature tooling. Yet, the widespread deployment of many refractory metallic systems, particularly ones with body-centered cubic (BCC) crystal structures, is fundamentally constrained by brittleness at low temperatures~\cite{miracle2024strength,mrotzek2006hardening}. A pronounced brittle-to-ductile transition temperature (DBTT) demarcates the regime below which tensile ductility collapses and cleavage-dominated fracture prevails~\cite{naujoks1996tungsten,abernethy2017predicting}. This transition imposes severe limitations during processing treatments and low-temperature service conditions, where embrittlement can compromise structural integrity and reliability~\cite{krimpalis2017comparative,terentyev2017mechanical,ren2018methods,Kim2014}. Importantly, the DBTT is a highly sensitive function of the microstructural state, wherein variables including processing pathways, impurity content, surface condition, strain rate, and microstructural length scales exert profound control over the transition temperature and, by extension, the macroscopic mechanical properties~\cite{kumar2024degradation,cook2024kink,geach1955alloys,gumbsch1998controlling,ren2018methods,farrell1967recrystallization}. Understanding and engineering this microstructural sensitivity of the DBTT remains central to unlocking the full potential of refractory BCC metals for next-generation extreme-environment applications, and requires an atomistic-level understanding of the failure mechanisms and their coupling to the evolving defect microstructure.


Tungsten provides a particularly useful model system for examining these questions because it is a prototypical refractory BCC metal whose brittle behavior is known to depend strongly on microstructure~\cite{gumbsch1998controlling,butler2018mechanisms}. A variety of deformation mechanisms have been observed in tungsten depending on the defect state, length scale, loading conditions, strain rate, and temperature~\cite{dummer1998effect,srivastava2013dislocation,wang2015situ,wang2018consecutive,wang2020anti,wang2022discrete}. At lower temperatures, coarse-grained polycrystalline tungsten fractures primarily along grain boundaries with little or no plastic deformation~\cite{talignani2022review}. Despite this apparent brittleness, the ductility of tungsten can be improved substantially through thermomechanical processing. For example, ultrafine-grained, dislocation-rich tungsten produced by severe cold and warm rolling exhibits significant room-temperature ductility~\cite{reiser2016ductilisation,bonnekoh2018brittle,bonnekoh2020brittle,lied2021comparison,ren2019investigation}. Conversely, recrystallized or defect-poor polycrystalline tungsten becomes significantly more brittle~\cite{farrell1967recrystallization,shah2020brittle,richou2020recrystallization,miracle2024strength}. At the same time, experiments on large macroscopic, high-quality single crystals further suggest that tungsten is not intrinsically brittle. These tensile tests show that single crystals can sustain substantial plastic strains of up to 20\% at temperatures as low as 140 K, with necking occurring prior to fracture in all cases down to 20 K when the surface finish is carefully controlled~\cite{beardmore1965deformation}. At the small-scale extreme, \textit{in situ} TEM studies have demonstrated that twinning can dominate ductile tensile deformation in tungsten nanopillars loaded along the $\langle100\rangle$, $\langle110\rangle$, and $\langle111\rangle$ directions~\cite{liu2014atomistic,wang2015situ,wang2020unstable,zhong2024atomic}, while analogous compression experiments suppress twinning, underscoring the strong role of stress state~\cite{kim2010tensile}. Deformation conditions can significantly affect the competition between dislocation slip and twinning, and atomic-scale observations have revealed processes such as twin growth and self-detwinning mediated by motion of incoherent boundary segments~\cite{wei2019bending}. Multiple studies have directly connected twins to fracture behavior. It was shown that twins can nucleate cracks and that propagating cracks can in turn generate twins in BCC films~\cite{liu2014atomistic}. In polycrystalline tungsten subjected to high strain-rate loading, fracture has been observed to initiate at twin–twin and twin–grain-boundary intersections~\cite{dummer1998effect}, while in nanoscale specimens twin boundaries reduced fracture strength by serving as preferred crack initiation sites~\cite{cui2023effect}. Other studies have suggested that twin-mediated crack nucleation plays an important role in low-temperature fracture of tungsten single crystals and shock-loaded material~\cite{argon1966fracture,asay1980shear}.

Taken together, these observations point to a general picture in which multiple strain-accommodating mechanisms, including dislocation slip, twinning, and fracture compete locally during deformation. Which mechanism is activated first depends on the local microstructural state, including the density and arrangement of defects, and the local stresses generated by their interactions, which together determine the activation barriers for each defect process. Understanding how these coupled processes evolve and interact across length scales is therefore essential for identifying the microscopic pathways that lead to brittle fracture.

While experimentally resolving how different types of deformation mechanisms interact during deformation remains a formidable challenge, recent advances in high-performance computing now enable atomistic simulations at unprecedented scales, allowing us to model sample volumes that are large enough to host statistically meaningful defect microstructures while retaining full atomistic resolution of every deformation event.~\cite{zepeda2017probing} This cross-scale molecular dynamics approach allows us to follow, within a single unified framework, the collective evolution of realistic dislocation networks together with any new defects that can emerge during deformation, including deformation twins and cracks, defects that cannot be simultaneously resolved with any other available method. 
This cross-scale approach has been successfully applied in recent years to address several long-standing scientific questions, including single-crystal plasticity and steady-flow regimes~\cite{zepeda2017probing}, the mechanisms of stage hardening and crystal-orientation attractors~\cite{zepeda2021atomistic,bertin2023crystal}, metallic alloys~\cite{islam2025nonequilibrium}, and the tension/compression asymmetry in BCC metals~\cite{bertin2022sweep}. Previous cross-scale MD studies focused exclusively on single crystals with periodic boundary conditions, predicting fully ductile behavior of refractory metals even at extreme strain rates and room temperature, seemingly contradicting experimental observations. On the other hand, traditional MD simulations, which included free surfaces were limited to nanoscale volumes and predicted unrealistic failure mechanisms near the ideal strength of a perfect crystal lattice~\cite{cui2023effect}. In this work, we demonstrate the first application of extreme-scale MD to fracture in refractory metals, incorporating realistic surfaces and dislocation microstructures to reveal fracture emerging from tightly coupled interactions among different defects including dislocations, disconnections and twin boundaries spanning atomic-to-micrometer scales. In combination with TEM observations, this approach enables us to identify a crack-nucleation mechanism mediated by twin-boundary pinning and disconnection pile-up, relate this mechanism to the DBTT, and clarify how stored dislocations can delay starvation and promote ductility.

\section{Results}
\label{sec:Results}

\subsection{Effect of defect microstructure on ductility and brittleness of tungsten}

\begin{figure}[th!]
\begin{center}
     \begin{tabular}{c}
  \includegraphics[width=\textwidth]{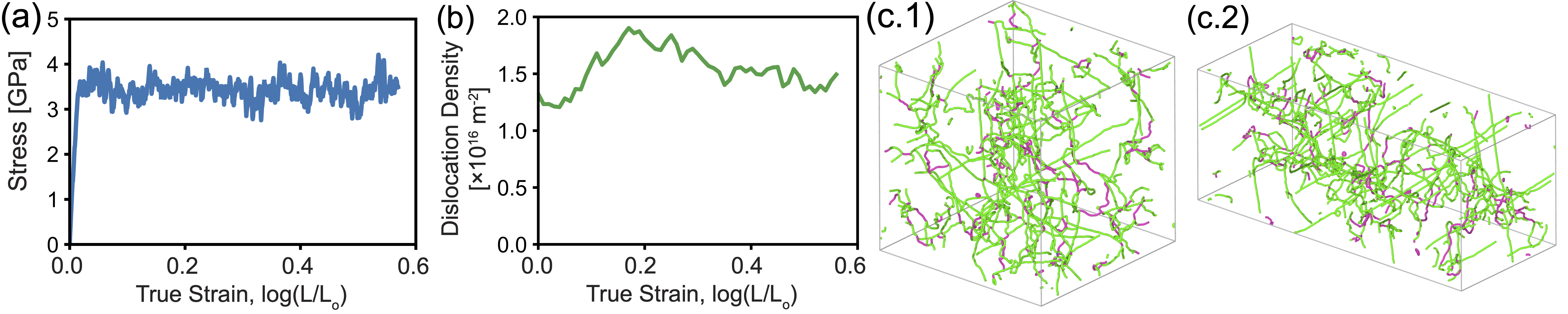}
\end{tabular}
\end{center}
  \caption{\label{fig:Fig2} Ductile behavior at lower temperatures and high straining rates was observed in simulations with periodic boundary conditions initiated with dislocation networks. The blocks can be deformed to essentially arbitrary strains (c) while maintaining constant dislocation density (b) and flow stress (a).}
\end{figure}

\begin{figure}[th!]
\begin{center}
     \begin{tabular}{c}
  \includegraphics[width=\textwidth]{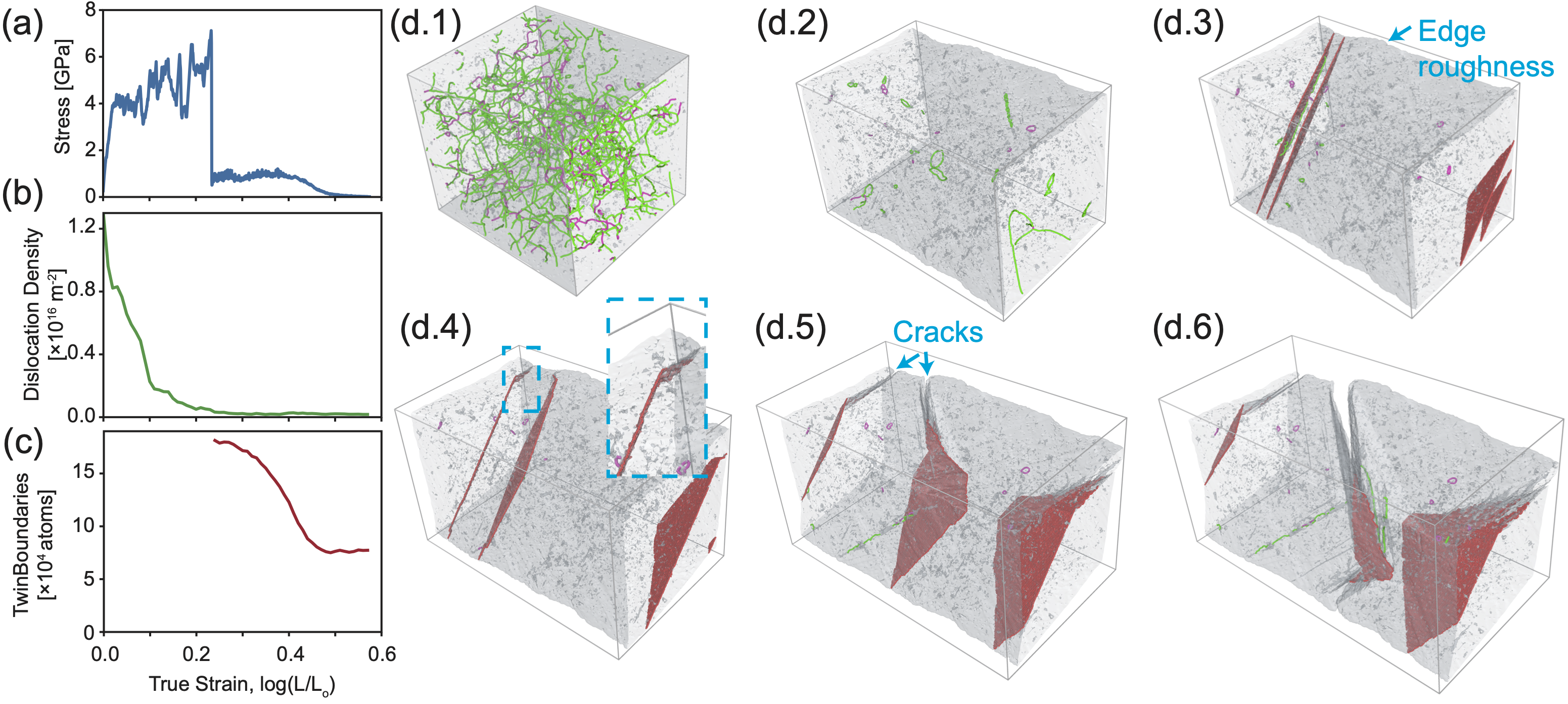}
\end{tabular}
\end{center}
  \caption{\label{fig:Fig3} Brittle behavior was observed at low temperatures in simulations with surfaces. The initial dislocation starvation stage (d.1,d.2) leads to increasing stress (a) and twin nucleation (d.3). While the twin growth initially results in a ductile deformation, eventually the moving twin boundaries get pinned by surface imperfections resulting in a disconnection pile-up near the surface pinning site. This pile-up leads to the formation of an inclined, incoherent twin boundary segment, which serves as a critical site for crack nucleation and propagation (d.4-d.6) at low macroscopic stress (a).}
\end{figure}

We begin by demonstrating that tungsten samples with pre-existing dislocation networks deform in a ductile manner under periodic boundary conditions but exhibit brittle fracture when open surfaces are introduced.
First, we performed $\langle100\rangle$ tensile deformation simulations using fully periodic boundary conditions at 300~K with a deformation rate of $10^7$ s$^{-1}$. The sample deformed in a ductile manner throughout the entire strain range of $\sim0.6$. Figures~\ref{fig:Fig2}(a) and \ref{fig:Fig2}(b) show, respectively, the stress and the dislocation density as a function of true strain. The flow stress remained nearly constant at an average value of 3.5 GPa, while the steady-state dislocation density fluctuated around $1.5 \times 10^{16}$ m$^{-2}$. While tungsten’s poor ductility is often attributed to the low mobility of screw dislocations, our simulations show that when a dislocation network is present, the dislocations remain sufficiently mobile, allowing this EAM model of W to exhibit ductile behavior even at high deformation rates. The observed plasticity is consistent with previous MD studies in W and Ta~\cite{bertin2022sweep, bertin2023crystal}.

In contrast, brittle fracture shown in Fig.~\ref{fig:Fig3} was observed when free surfaces were introduced along the lateral directions. The deformation process occurs in three distinct stages, each characterized by changes in the type of deformation defects, as also reflected in the stress-strain curve, Figs.~\ref{fig:Fig3}(a, b, c). The snapshots in Figs.~\ref{fig:Fig3}(d.1) to \ref{fig:Fig3}(d.6) illustrate the evolution of the defect microstructure during the simulation, ultimately leading to crack nucleation and failure.  

The first stage of the deformation is ductile and exhibits dislocation network-mediated plasticity. During this stage, the stress-strain curve indicates strain hardening. Beyond yield, the gradual flow stress increase is commensurate with the rapid decrease of the dislocation density resulting from dislocations escaping to the free surfaces, Figs.~\ref{fig:Fig3}(a and b). At a critical strain of about $\sim 0.2$, complete dislocation starvation occurs~\cite{brenner1957plastic,brenner1956tensile,greer2005size,greer2006nanoscale,greer2009situ,sharma2018nickel,frick2008size,uchic2004sample}; with no mobile defects left in the crystal to carry plasticity, the stress rises rapidly as a result of the continued applied deformation. Interestingly, the observed strain hardening behavior is the opposite of what is expected from the well-known Taylor's law that relates flow stress to dislocation density in bulk~\cite{taylor1934mechanism}. Previous cross-scale simulations with periodic boundary conditions linked hardening to crystal rotation~\cite{zepeda2021atomistic}. However, our simulation shows no measurable rotation, and the crystal maintains its ⟨100⟩ orientation parallel to the loading direction throughout. Prior experimental studies have also reported strain hardening under conditions of dislocation starvation, concluding that the progressive exhaustion of mobile dislocations and dislocation sources increases the stress required to sustain plastic flow~\cite{chisholm2012dislocation,shan2008mechanical}. Our simulations are consistent with these experimental observations of exhaustion hardening and further suggest that dislocation starvation progressively eliminates segments of the dislocation network that operate at lower stresses resulting in counterintuitive strengthening.

The second stage begins when the stress reaches the nucleation threshold of a single twin at approximately 6 GPa. Immediately after nucleation, the twin begins to grow as  shown in Figs.~\ref{fig:Fig3}(d.3), and the flow stress drops significantly to around 1.5 GPa, Fig.~\ref{fig:Fig3}(a). This  behavior is consistent with prior \textit{in situ} high-resolution TEM observations of ductile deformation by twinning in BCC W pillars ~\cite{wang2015situ}. During the twinning stage, twin growth is accompanied by coupled motion of the two coherent \(\Sigma 3(112)[110]\) twin boundaries (TBs) shown in Figs.~\ref{fig:Fig3}(d.3)~\cite{cahn2006coupling}. Their motion is mediated by the nucleation and subsequent migration of individual disconnections~\cite{han2018grain}, which nucleate at the TB-surface triple junctions on one side of the deforming sample. We observed two distinct structures or phases of the twin~\cite{faisal2021modeling}, separated by a grain boundary phase junction~\cite{frolov2021dislocation}, with the motion of these grain boundary phase junctions contributing to TB migration. It has been shown recently that the Burgers vector of these defects is necessarily smaller than that of regular disconnections~\cite{winter2025quantifying}, which is likely the reason they get activated leading to TB phase transformation-mediated migration. The relatively low flow stress of this stage shown in Fig.~\ref{fig:Fig3}(a) is determined by the nucleation and motion stress of these TB line defects.

The third stage of deformation is marked by (i) the pinning of the twin boundary at one of the sample’s edges due to surface roughness, (ii) the formation of an inclined incoherent TB segment, and (iii) crack nucleation. Fig.~\ref{fig:Fig3}(d.4) illustrates a moving TB as it becomes partially pinned at one end by surface roughness. During this process, disconnections continue to nucleate and propagate along the planar section of the TB; however, they are no longer able to escape on the opposite side, leading to a localized accumulation of disconnections (i.e., a ``pile-up'') on adjacent planes and ultimately results in the formation of an asymmetric, incoherent TB segment. After the asymmetric segment reaches a length of approximately 10 nm, we observe a formation of a brittle crack. The crack only propagates through the incoherent boundary segments. The subsequent growth of the crack is enabled by the continued migration of the coherent TB segment which stays intact and provides new disconnections/incoherent segments to be consumed by the advancing crack tip. Eventually, crack growth leads to the complete failure of the specimen at the end of the simulation, as shown in Fig.~\ref{fig:Fig3}(d.6). Note that the sample fails by this mechanism at relatively low macroscopic stress around 1.5 GPa, which is significantly lower than bulk flow stress of ~4 GPa or the twin nucleation stress of ~7 GPa.

\begin{figure}[th!]
\begin{center}
     \begin{tabular}{c}
  \includegraphics[width=\textwidth]{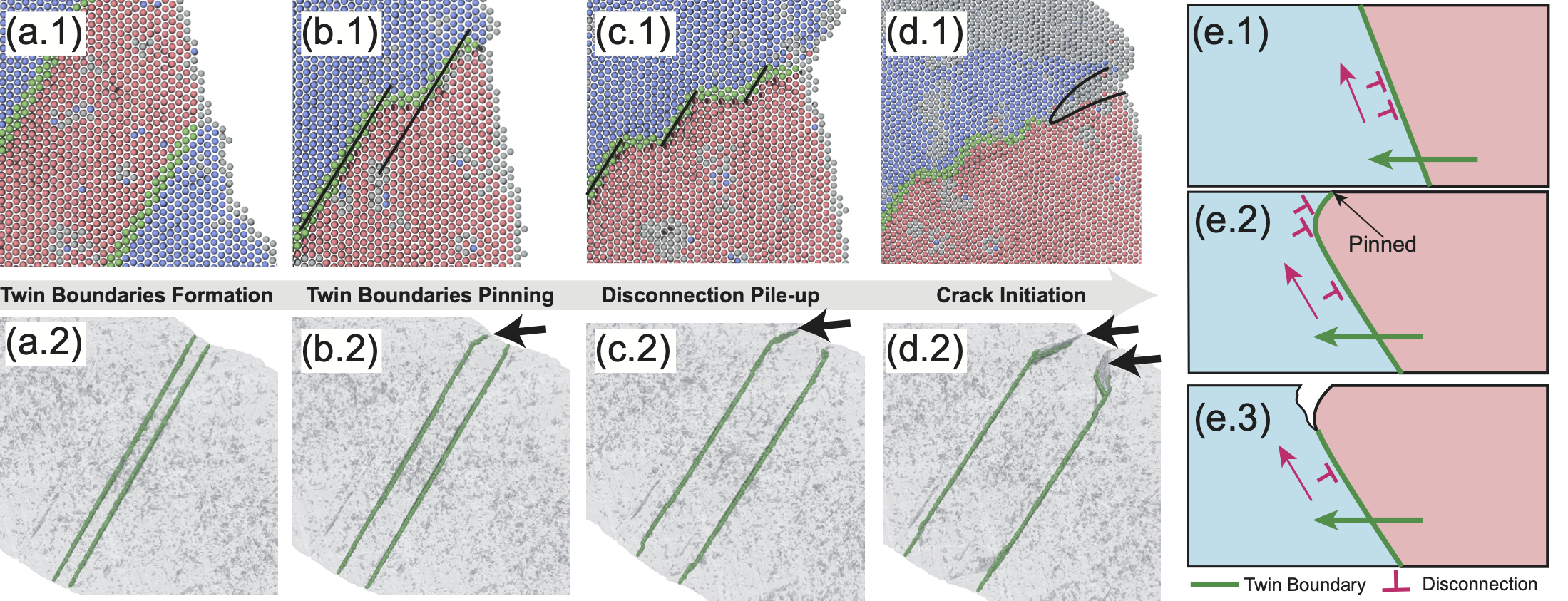}
\end{tabular}
\end{center}
  \caption{\label{fig:Fig4} Crack nucleation mechanism by disconnection pile-up at twin boundaries. (a) A twin formed by two planar coherent twin boundaries that migrate during twin growth by nucleation and propagation of disconnections. Snapshots (a.1) and (a.2) show the same atomistic TB configuration viewed at different scales. (b) Early stage of twin boundary pinning with a single disconnection stuck near the surface roughness. (c) Continuous TB migration results in a disconnection pile-up and the formation of a longer inclined boundary segment. (d) Crack nucleation and propagation through the disconnection pile-up region. (e) Schematic illustrating the disconnection pile-up mechanism.}
\end{figure}

The atomistic details of the disconnection pile-up leading to crack formation are illustrated in Figure \ref{fig:Fig4}. Figures \ref{fig:Fig4}(a.1) and \ref{fig:Fig4}(a.2) show atomic and specimen scale views of the deformation twin, which is initially bounded by two flat TBs. Figure~\ref{fig:Fig4}(b.1) presents an atomic scale view of a single disconnection with its characteristic step on the TB. The disconnection becomes pinned at the surface, rendering itself immobile. At the specimen scale, Fig.~\ref{fig:Fig4}(b.2), this immobile disconnection leads to the formation of an inclined boundary plane. As more disconnections move through the flat portion of the TB but are unable to escape, a disconnection pile-up forms, as depicted in Fig.~\ref{fig:Fig4}(c.1). This accumulation results in the extension of the inclined boundary plane, as shown in Fig.~\ref{fig:Fig4}(c.2). Finally, Fig.~\ref{fig:Fig4}(d) illustrates the nucleation and growth of a crack in the region near the disconnection pile-up. The fracture mechanism is also schematically represented in Fig.~\ref{fig:Fig4}(e). These simulations clearly demonstrate that the same material model can exhibit both ductile and brittle behavior under identical straining and temperature conditions depending on the defect microstructure.

\subsection{High-temperature simulations of ductile-to-brittle transition}
We now proceed to investigate the effect of temperature and demonstrate that our model material exhibits a ductile-to-brittle transition. Standard fracture mechanics assumes that a material will fail in ductile fashion when dislocations near the crack tip can move, effectively allowing for mechanical energy to be dissipated via plastic slip. In most BCC metals, with plastic flow controlled by screw dislocations, temperature reduces the critical dislocation glide stress, allowing for ductile crack propagation. This thermal softening is generally not gradual, but sudden, which leads to a clear temperature transition between brittle and ductile regimes, the so-called DBTT. However, fracture mechanics is predicated on the intrinsic existence of subcritical cracks in bulk specimens. When this premise is not satisfied, the DBTT is controlled by other mechanisms that depend on the material's defect microstructure.
Next, we investigate whether the brittle behavior observed in the simulations with free surfaces leads to ductile conditions at high temperatures.

\begin{figure}[th!]
\begin{center}
     \begin{tabular}{c}
  \includegraphics[width=0.8\textwidth]{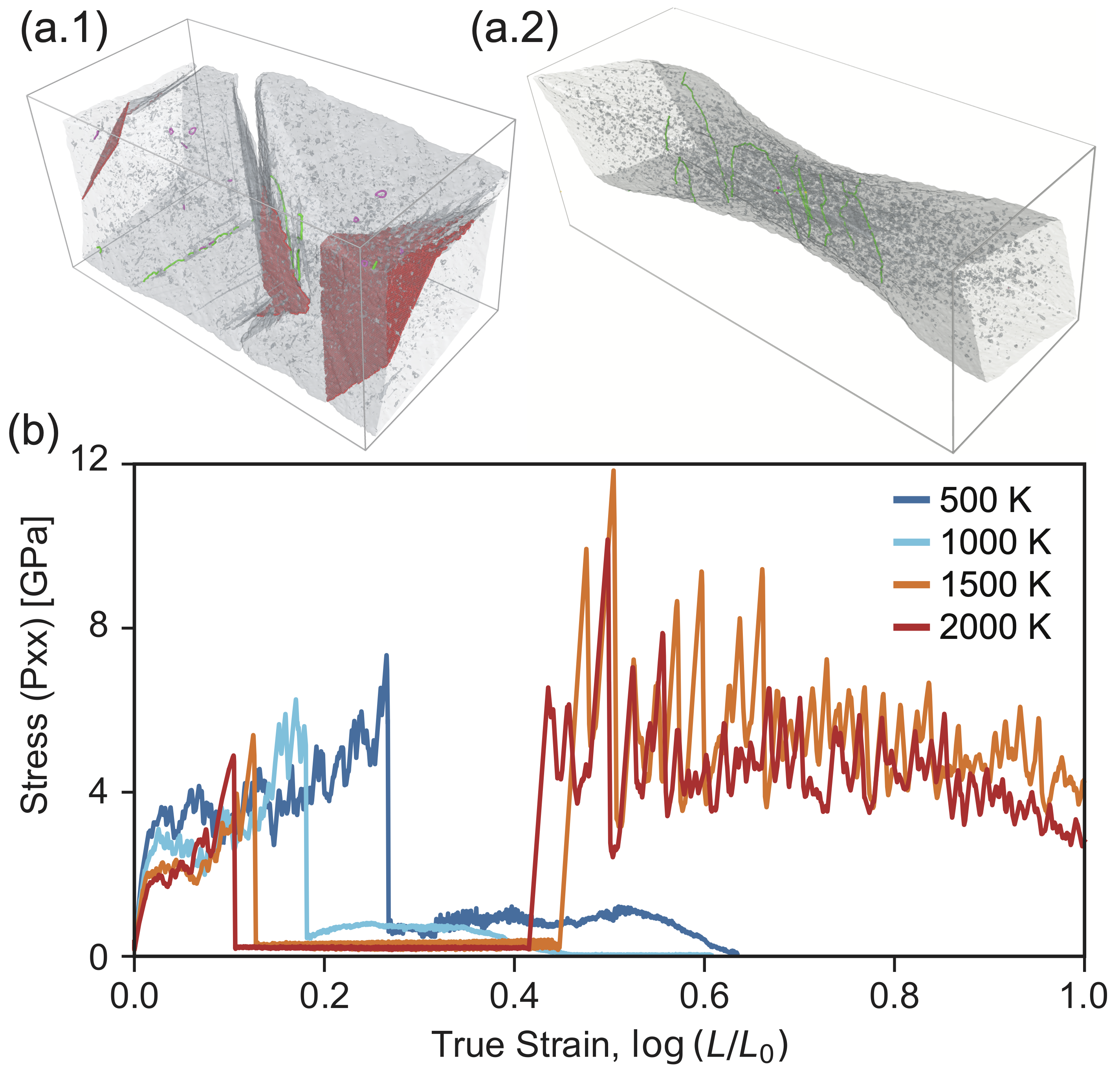}
\end{tabular}
\end{center}
  \caption{\label{fig:Fig6} The ductile-to-brittle transition temperature (DBTT) is predicted by simulations with pre-existing dislocations and free surfaces. (a) The DBTT is marked by the absence of twin boundary pinning and disconnection pile-up. (b) The stress-strain curves show brittle fracture at low temperatures and ductile behavior at high temperatures. While the two initial stages, including dislocation starvation and twin nucleation, occur at all temperatures, high-temperature samples continue to deform in a ductile manner through dislocation nucleation after twin boundaries migrate through the sample without pinning and eventually annihilate.}
\end{figure}

Simulation results for temperatures ranging from 500 to 2000~K are shown in Figure \ref{fig:Fig6}. The stress-strain curves for all simulations are shown in Fig.~\ref{fig:Fig6}(b). The first two stages of deformation are common to all simulations: dislocation starvation characterized by an increase in the flow stress eventually followed by twin nucleation and  sudden drop in stress. At $T\leq1000$ K, the two initial stages are followed by fracture via the crack nucleation mechanism identified in Fig.~\ref{fig:Fig4}, whereas starting at 1500 K and above, no cracks are observed. This is suggestive of a DBTT in our model sample between 1000 and 1500~K. 
Remarkably, below the DBTT, the samples are seen to fracture at smaller strains as temperature increases. For example, at 500~K the samples fail at a strain of $\sim$0.62, whereas at 1000~K fracture occurs at strain $\sim$0.43. This counterintuitive behavior follows directly from the dislocation starvation mechanism: at higher temperatures, increased dislocation mobility leads to faster starvation, causing failure to occur at progressively smaller strains. Thus our simulations show an example where, below the DBTT, increasing temperature can make the material more brittle.

\begin{figure}[th!]
\begin{center}
     \begin{tabular}{c}
  \includegraphics[width=\textwidth]{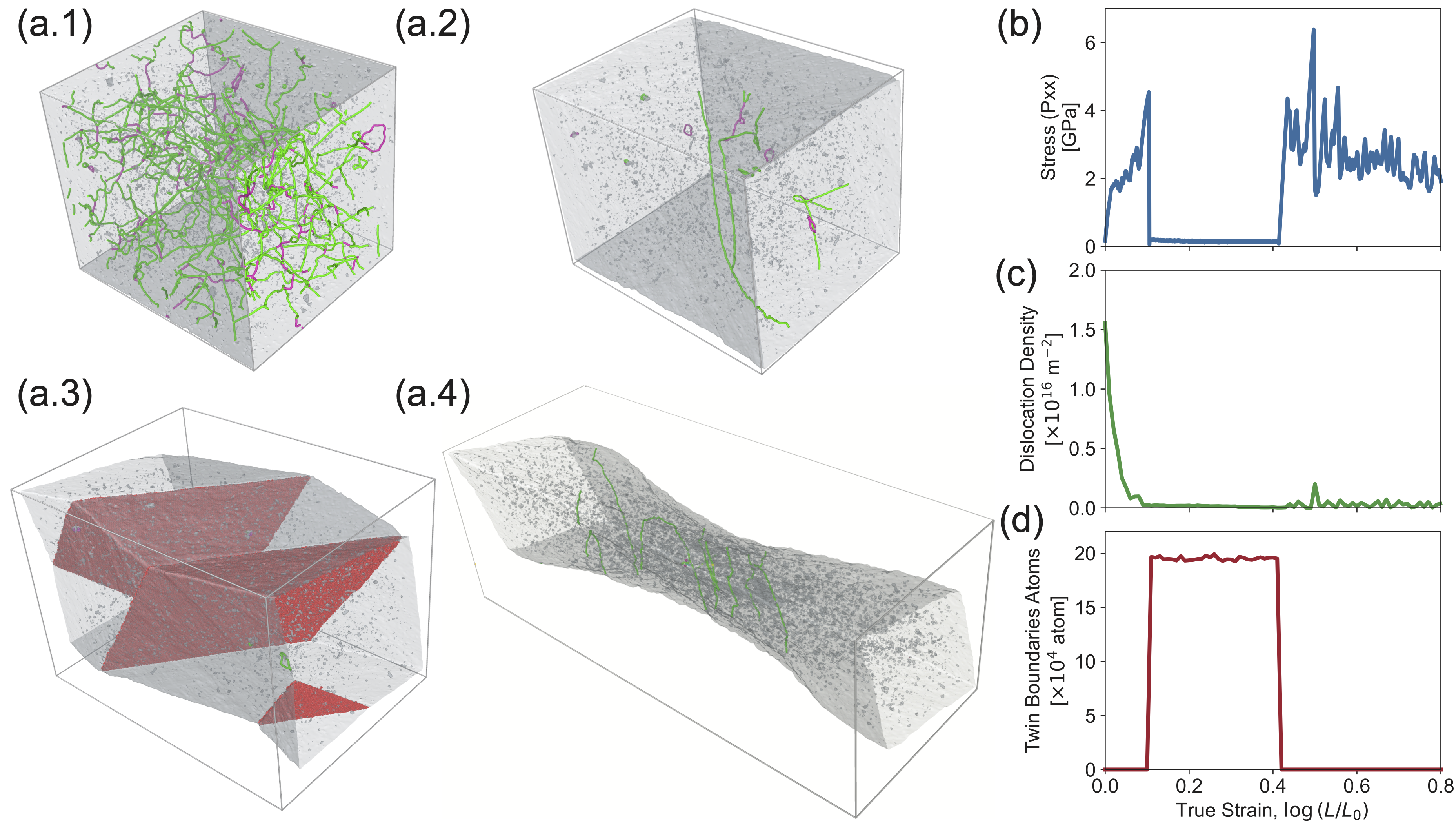}
\end{tabular}
\end{center}
  \caption{\label{fig:Fig5} Ductile deformation of samples with lateral free surfaces at high temperatures. Although the deformation proceeds through the same stages of starvation (a2) and twinning (a3), at higher temperatures twins do not get pinned by surface defects and no fracture occurs. Twin propagation proceeds in a ductile manner until the two TBs annihilate with each other across the periodic boundary conditions. The subsequent deformation proceeds by nucleation of dislocations at surfaces which eventually leads to gradual necking (a4). (b-d) Stress, dislocation density, and TBs atoms as a function of strain  reflecting the three different deformation stages. The samples were deformed at a strain rate of $10^7$ s$^{-1}$ at 2000~K.}
\end{figure}

Above the DBTT, the specimen displays features indicative of ductile behavior. These tests share the first two stages of deformation with the brittle samples, namely, dislocation starvation followed by twin nucleation. However, in contrast to the low temperature cases, the twin boundaries are seen to remain mobile and avoid pinning upon contact with the surface. The process is illustrated in Figure~\ref{fig:Fig5} for the representative case of 2000~K. Figs.~\ref{fig:Fig5}(a.1), \ref{fig:Fig5}(a.2), and \ref{fig:Fig5}(a.3) show, respectively, the initial configuration, the dislocation-starved configurations, and fully-formed twins. With the twin boundaries free to move along the free surface, the twinned region continues to grow until the entire cell is traversed by the TBs across the periodic boundaries. This results in fully twinned $\langle110\rangle$-oriented crystal containing no dislocations and a subsequent sharp rise of the flow stress, shown in Fig.~\ref{fig:Fig5}(b). 
The subsequent deformation proceeds in the starvation regime via nucleation of dislocations at free surfaces, indicating that twin nucleation is less favorable than dislocation nucleation in this new orientation. Continuation of this process leads to the gradual necking of the sample, Fig.\ref{fig:Fig5}(a.4).
Fig.~\ref{fig:Fig5}(b-d) shows the corresponding stress, dislocation density, and
TBs atoms plots reflecting the three distinct deformation stages.

The absence of fracture in our high-temperature simulations is directly correlated to the absence of TB pinning. Indeed, it is reasonable to expect that TB disconnections become more mobile at high temperatures and escape to the surface, preventing the disconnection pile-up that leads to fracture at low temperatures. In addition to the increased disconnection mobility, the extent of surface roughness also plays a role. For this, we examined the roughness of the surfaces immediately after complete starvation in simulations below and above the DBTT and observed that the high-temperature surfaces are noticeably smoother. Indeed, after the deformation in the ductile stage, a significant amount of surface steps are expected to be created as dislocations exit the material to open surfaces. Apparently, much of the pre-existing dislocation density also annihilates inside the crystal at high temperatures, resulting in fewer dislocations to cross the free surfaces, hence producing less surface steps. Smooth surfaces result in less pinning of TBs, thereby preventing fracture. 

The observed low-temperature ductile deformation by twinning also suggests that the DBTT of our model could be lowered significantly by removing or polishing the surface roughness. In contrast, introducing large obstacles on the surface could lead to brittle fracture even at high temperatures. We argue that, at least in our model, DBTT is not an inherent materials property dictated by dislocation properties, but instead depends on the materials microstructure. This conclusion is consistent with prior experimental studies that reported a wide range of DBTTs for differently processed materials~\cite{xue2022brittle}.

\subsection{Stored dislocations to enhance ductility}

\begin{figure}[th!]
\begin{center}
     \begin{tabular}{c}
  \includegraphics[width=0.5\textwidth]{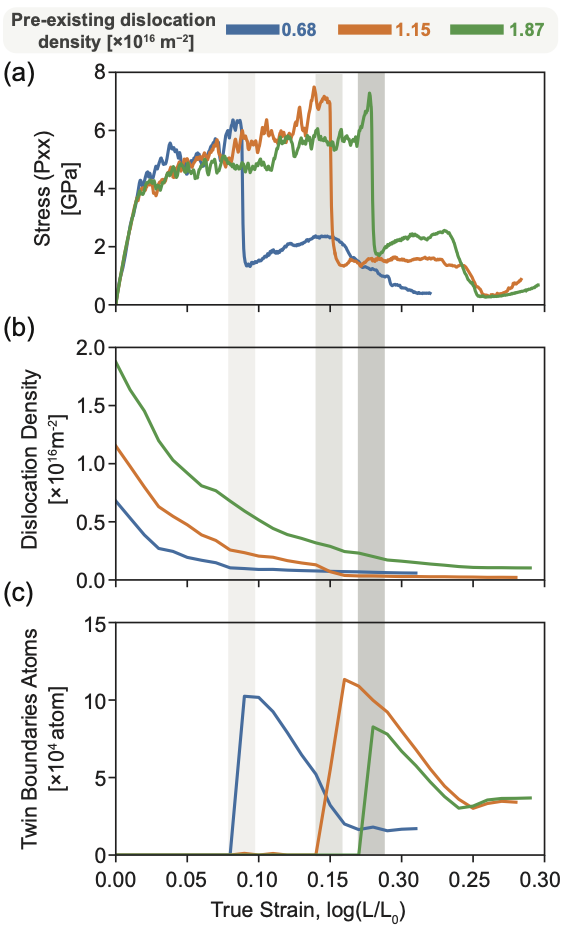}
\end{tabular}
\end{center}
  \caption{\label{fig:Fig7} Larger initial density of stored dislocations delays starvation and fracture. (a) Stress, (b) dislocation density and (c) TBs atoms as a function of true strain for different pre-existing dislocation densities at a strain rate of $10^7$~s$^{-1}$ and temperature of 300~K.}
\end{figure}

In this section, we return to simulations at the ambient temperature of 300~K and investigate the effect of stored dislocations, such as those generated by thermo-mechanical treatment, on the sample’s ductility. Our simulation results demonstrate that a higher density of pre-existing dislocations delays fracture. We prepared three different computational samples with different dislocation densities using a pre-compression stage described in the {\it Methods} section. The samples were then deformed in tension along the $Z$ direction at a strain rate of $10^7$ s$^{-1}$ with free surfaces along the lateral directions.
%
Figure~\ref{fig:Fig7} illustrates the stress-strain curves along with the evolution of dislocation density and TB during deformation. All three simulations produce self-similar curves with three characteristic stages: dislocation starvation and exhaustion hardening, deformation by twinning, and finally, fracture.

All three samples yield at very similar stress despite the noticeably different dislocation densities. After the yield point, we observe exhaustion hardening manifested by the increasing flow stress. During the hardening stage, the dislocation density quickly decreases because dislocations escape to free surfaces. As noted before, this behavior is opposite of what is expected from the well-known Taylor’s hardening law that describes bulk behavior.

As was shown earlier for this type of initial defect microstructure and boundary conditions, the hardening stage and increasing flow stress continue until all the dislocations are exhausted by exiting through the free surfaces. Subsequently, twin nucleation and deformation by twin growth takes place which eventually leads to fracture. Thus, by increasing the initial dislocation density, dislocation starvation is delayed, thereby extending the initial ductile stage. 

\subsection{Twinning along cracks in bulk tungsten}
\begin{figure}[th!]
\begin{center}
     \begin{tabular}{c}
  \includegraphics[width=\textwidth]{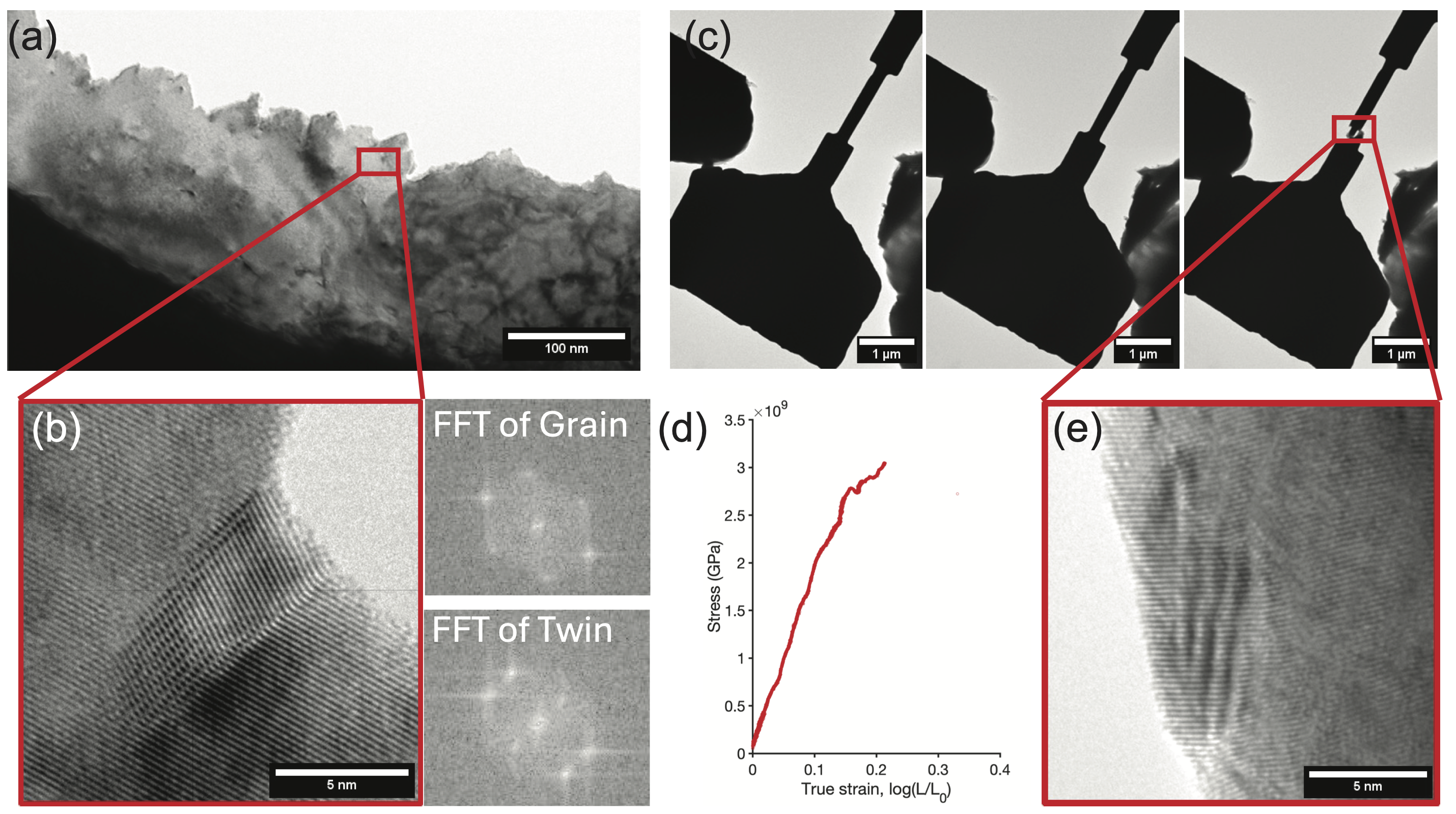}
\end{tabular}
\end{center}
  \caption{\label{fig:bulk_experiment}(a) TEM image showing the edge of the crack path, and (b) a higher-magnification image highlighting an incoherent twin pocket and the corresponding fast Fourier transforms of the grain and twin. Such incoherent twins are frequently observed along the fracture path. (c) Time-lapse TEM images from an in-situ tensile test and (d) the corresponding stress–strain curve. (e) Post-fracture image showing an incoherent twin at the fracture surface, with an inset displaying its fast Fourier transform.}
\end{figure}

The simulations above point to a twin-dependent mechanism for crack initiation and propagation in W.  Nanoscale W specimens loaded in tension exhibit twinning consistent with our MD results~\cite{zhong2024atomic}.  Evidence for twinning in bulk W at room temperature, however, remains sparse and indirect~\cite{argon1966fracture}. The MD simulations indicate that twin nucleation occurs at high stresses, which might only be observed in dislocation starved microstructures or near singular points like surface flaws or crack tips. To examine the prevalence of twinning in bulk specimens, 
25-$\mu$m-thick coarse grained tungsten samples were fractured under tensile loading and directly loaded into the transmission electron microscope.  Much of the crack path is not visible, but certain prominent features protrude sufficiently beyond the surface to observe the crack path directly without any additional sample preparation. Small 'pocket-shaped' twinned regions lay along the edge of the fracture surface, see example in Figure~\ref{fig:bulk_experiment}. Incoherent twin boundaries primarily surround these twin pockets, which exhibit mostly curved segments. These observations align with prior \textit{in situ} TEM results on ~15 nm nanowires that observed the formation of similar twin geometries during loading and unloading. To characterize the stress associated with twinning and observe their formation at intermediate length scales  micro-tensile specimen were prepared by focused ion beam milling. The samples were loaded to failure \textit{in situ} in the TEM. The example in Fig.~\ref{fig:bulk_experiment} fractured at ~3 GPa. \textit{In situ} post-mortem analysis of the fracture surface reveals the presence of similar twin pockets at the fracture surface. The multiscale observations of nanoscale twin pockets demonstrate their ubiquitousness at tungsten fracture surfaces. These geometries differ from the parallel twins with large coherent segments simulated above.  The experiments, however, characterize random crystal orientations.  


\begin{figure}[th!]
\begin{center}
     \begin{tabular}{c}
  \includegraphics[width=\textwidth]{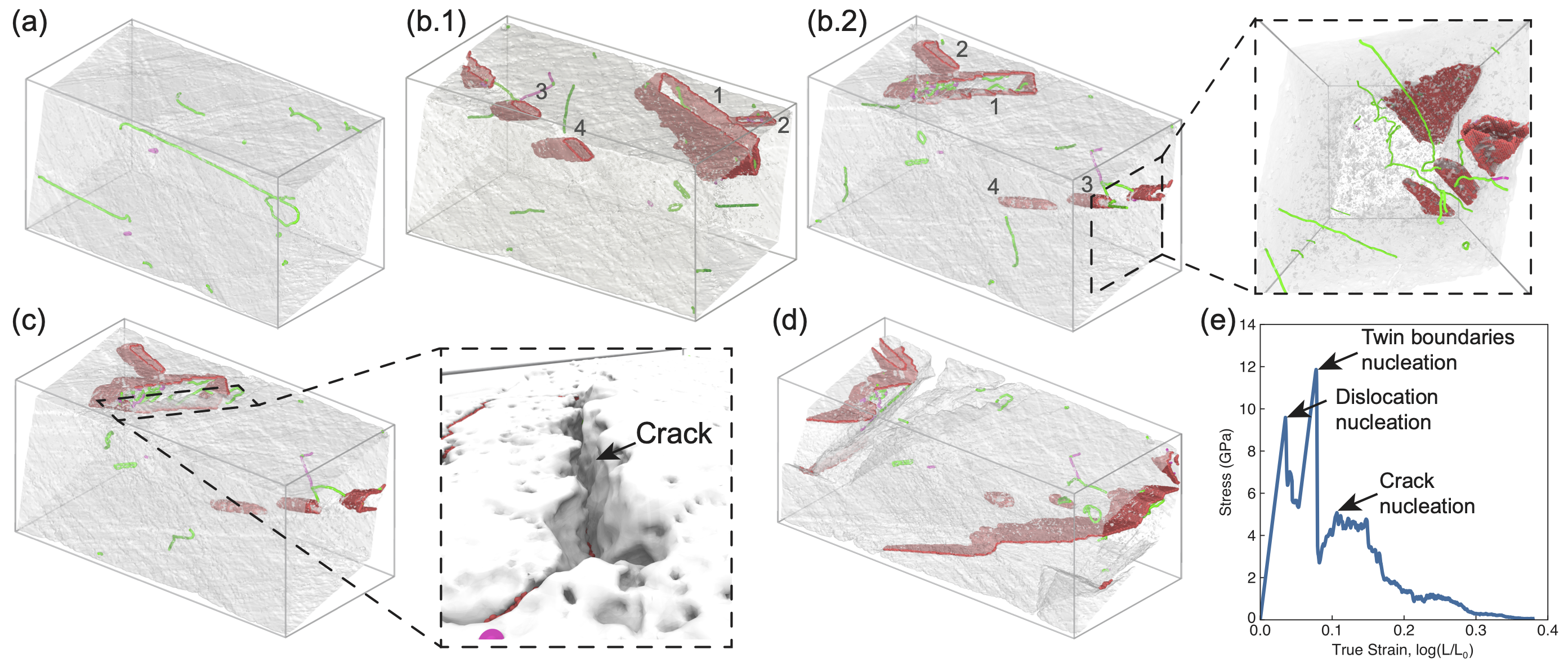}
\end{tabular}
\end{center}
  \caption{\label{fig:figure110}Twin-pocket formation on strained crystal surfaces as precursors to crack nucleation along incoherent TB segments shows the same fracture mechanism for other orientations. (a) Pre-starved crystal loaded along $\langle110\rangle$ direction initially deforms by nucleation of dislocations on surfaces (b.1) and (b.2) Subsequent nucleation of four twin pockets, surface grains separated from the main crystal by TBs, shown from two different perspectives.
  (c) Crack nucleation occurs along an incoherent TB segment of twin pocket 1. (d) Propagation of the crack leading to complete crystal fracture. (e) Stress-strain curve indicating activation stresses for each of the three events, with crack nucleation occurring at the lowest stress.  }
\end{figure}

\subsection{Effect of crystal orientation}

\begin{figure}[th!]
\begin{center}
     \begin{tabular}{c}
  \includegraphics[width=\textwidth]{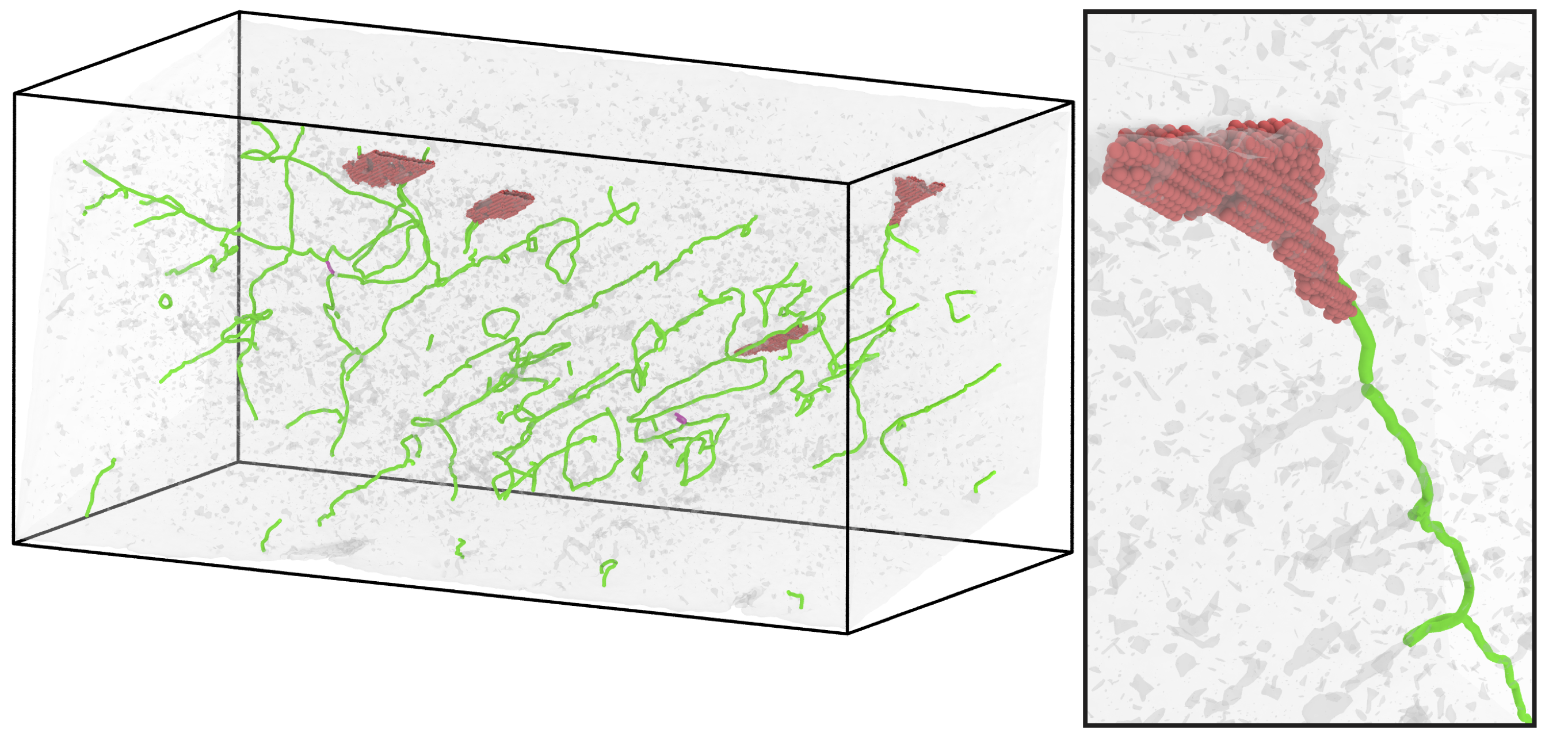}
\end{tabular}
\end{center}
  \caption{\label{fig:Tulip_Fig} Subcritical twin embryo dynamically appearing and disappearing at surface--dislocation intersections during early stages of loading . The crystal is loaded in tension along the $[110]$ direction. The main panel shows the dislocation network (green lines) and emerging twin structures, while the zoom-in view highlights the ``tulip-like'' morphology of the nucleated twin embryo (red).}
\end{figure}

The simulations discussed thus far pertain to initially $\langle100\rangle$-oriented crystals and clearly demonstrated a disconnection-mediated fracture mechanism along incoherent TBs. To make a better connection with the experimental results described above, we now show that other orientations, less prone to twinning, exhibit the same experimentally observed twin pockets that fracture along their incoherent TB segments, confirming the same underlying mechanism. To that end, we consider an annealed $\langle110\rangle$-oriented crystal in which almost no dislocations remain, as shown in Fig.~\ref{fig:figure110}(a).
Upon tensile loading, the sample first yields through the nucleation of surface dislocations, exhibiting a peak yield stress of approximately 9.8 GPa. During this stage, we observe the dynamic formation and disappearance of subcritical nuclei of twin pockets at dislocation–surface intersection sites, as shown in Fig.~\ref{fig:Tulip_Fig} (also see {\it Supplemental} Movie).

At a higher stress of about 12 GPa, supercritical twin pockets nucleate, labeled 1–4 in Fig.~\ref{fig:figure110}(b), and two distinct views (Fig.~\ref{fig:figure110}(b.1)–(b.2)) reveal their three-dimensional morphology. The inset in Fig.~\ref{fig:figure110}(b.2) highlights the faceted geometry of these surface grains, bounded by both coherent and incoherent TB segments. Following this event, the stress drops to approximately 4 GPa, and further deformation proceeds through continued dislocation activity and twin-pocket growth. When the incoherent TB segments reach a critical length, cracks nucleate along these interfaces (Fig.~\ref{fig:figure110}(c)). The crack then propagates along the incoherent segment, leading to sample fracture, as shown in Fig.~\ref{fig:figure110}(d). The corresponding stress–strain curve (Fig.~\ref{fig:figure110}(e)) captures these key microstructural transitions.

Notably, whereas the previously studied $\langle100\rangle$-oriented pillar fractured at a microscopic stress of only 1 GPa, the $\langle110\rangle$-oriented pillar cracked at 5.7 GPa, a value closer to the experimental 3.8 GPa. Although crystal orientation strongly influences the fracture stress, the underlying mechanism remains identical in both cases. The twin-pocket structures observed in our simulations closely match experimental observations, indicating that such twin pockets likely exist on the surfaces of crystals with various orientations and could serve as robust, common sites for crack nucleation.

\section{Discussion}
We investigated the ductility and fracture behavior of tungsten pillars as a function of initial defect microstructure, deformation conditions, and temperature using cross-scale molecular dynamics simulations and TEM observations. This approach enabled us to resolve, within a single framework, the coupled evolution of dislocations, twins, disconnections, and cracks in samples large enough to contain realistic defect microstructures. By bridging atomic-scale defect processes and near-micron-scale deformation, the simulations provide a mechanistic basis for the brittle and ductile responses of the model material.

Our simulations reveal that our tungsten model can exhibit both ductile and brittle behavior at low temperatures, depending on the defect microstructure. When a dislocation network is present in the crystal, the model material can deform at low temperatures and under high straining rates (on the order of $10^8$ s$^{-1}$) to essentially arbitrary strains in a ductile manner, suggesting that the dislocation mobility is not the limiting factor.
On the other hand, brittle fracture caused by disconnection pile-up at TBs was identified as a dominant failure mechanism in simulations with free surfaces. The simulations reveal a consistent sequence of events leading to failure. Deformation begins with the exhaustion of mobile dislocations: as dislocations escape to free surfaces, the crystal becomes increasingly hardened by starvation. To accommodate continued loading, the system nucleates coherent twin boundaries, which propagate until they encounter surface roughness. Pinning of these twin boundaries leads to the disconnection pile-up and formation of an inclined, incoherent segment. This region serves as the site for crack nucleation and subsequent propagation, allowing fracture to occur at macroscopic stresses far below slip mediated flow stress or twin nucleation stress. For other pillar orientations, we find a similar pathway, though preceded by the formation of surface twin pockets, which were also experimentally observed with high-resolution TEM in this study. These pockets grow and ultimately develop incoherent segments that again serve as the origin of fracture though at a higher macroscopic stress close to the experimental value. 
We also examined the role of stored dislocations and found that increasing their density delays starvation and extends the regime of ductile deformation, which is consistent with empirical observations of enhanced ductility of tungsten following hot rolling. This result supports the idea that processes reducing dislocation density, such as recrystallization, contribute to increased brittleness in tungsten.

The DBTT in our simulations was identified as the temperature above which TBs are no longer pinned by surface defects. The pillars fractured in all low-temperature simulations ($\leq$1000 K) as a result of disconnections accumulating at pinned TB segments, whereas at higher temperatures ($\geq$1500 K) the pillars exhibited ductile behavior including twinning and failed via progressive necking. Thus, the ductilizing effect of temperature in our simulations was not due to an increase in dislocation mobility, but rather to (i) smoother sample surfaces due to enhanced dislocation annihilation in the bulk at high temperatures, reducing TB pinning, and (ii) increased internal boundary and disconnection mobilities, allowing TBs to de-pin from obstacles more easily.

While the modeling predictions necessarily depend on the interatomic potential~\cite{hiremath2022effects}, the EAM model employed here reproduces a broad range of complex deformation behaviors previously observed experimentally, including dislocation starvation and exhaustion hardening~\cite{shan2008mechanical}, as well as twinning-dominated deformation in dislocation-starved pillars reported in \textit{in situ} TEM studies of W nanopillars~\cite{zhong2024atomic}. The simulations further capture crack nucleation at twin boundary–surface intersections, in direct qualitative agreement with fracture experiments on twinned nanopillars~\cite{cui2023effect}, high strain-rate compression tests of polycrystalline tungsten~\cite{dummer1998effect}, and our own TEM observations of twin pockets decorating deformed W surfaces. Taken together, these results suggest that despite its simplicity, the present potential provides a physically faithful representation of the coupled dislocation–twin–fracture processes governing brittle failure in tungsten, lending confidence in the mechanisms predicted by these simulations.

Finally, we discuss how the mechanisms identified here may contribute to brittle fracture in polycrystalline tungsten. Starvation and twinning followed by fracture may occur in small surface grains. Although deformation twinning is not generally considered a dominant mode in coarse-grained polycrystalline tungsten, twin boundaries, as the lowest-energy high-angle grain boundaries, should be an unavoidable component of any statistically representative polycrystalline microstructure. Under tensile loading, these twin boundaries will also experience shear stresses and driving forces for migration, and may become internally pinned by other boundaries, triple junctions or other defects, enabling disconnection accumulation and crack nucleation through mechanisms analogous to those identified here. 
More broadly, while experimental studies consistently show that fracture in polycrystalline tungsten occurs along high-angle grain boundaries,~\cite{dummer1998effect,wang2019cracking,talignani2022review} the microscopic mechanisms governing failure at these interfaces is not completely understood. In principle, fracture along general high-angle grain boundaries may also be mediated by boundary migration, disconnection pile-up, and pinning processes similar to those revealed in this study. Building on these insights, future work will apply cross-scale molecular dynamics to true polycrystals with free surfaces, examining how general high-angle grain boundaries behave under realistic combinations of shear, migration, and normal stress.

\section*{Methods} \label{sec:methods}

Molecular dynamics (MD) simulations of crystal plasticity are performed using the Large-scale Atomic/Molecular Massively Parallel Simulator (LAMMPS) software package~\cite{plimpton1995fast} and visualizations of atomistic structures are generated using OVITO~\cite{stukowski2009visualization}. Tungsten is modeled using an embedded atom method (EAM) interatomic potential by Zhou et al.~\cite{zhou2004misfit}, previously validated~\cite{xu2018deformation,maresca2020mechanistic} and employed to study the mechanical behavior of tungsten~\cite{mccoy2018structural,huang2017molecular,hu2023nanoindentation}. The OVITO implementation of the polyhedral template matching (PTM) algorithm~\cite{larsen2016robust} is used to identify the local structural environment of each atom, compute particle orientations, and reveal free surface and planar defects. The dislocation extraction algorithm (DXA)~\cite{stukowski2012automated} is used to identify dislocations within the atomistic system, and the BCC Defect Analysis (BDA)~\cite{moller2016bda} algorithm is used for precise identification of twin boundaries. 

Single-crystal BCC tungsten samples are generated within a box size of ($L_x, L_y, L_z$) = (200, 40, 40) nm containing about 20 million atoms. The crystal [100], [010] and [001] directions are chosen to align with the Cartesian frame $(X,Y,Z)$ of the simulated crystal. We employ simulation domains with various defect content in order to investigate the impact of the initial defect microstructure on the deformation behavior of tungsten. Specifically, we consider simulation domains with inexistent dislocations or pre-existing dislocation loops, geometries with periodic boundary conditions (PBCs) in all three directions, and domains with free surfaces in the lateral directions of the loading axis.

\begin{figure}[th!]
\begin{center}
     \begin{tabular}{c}
  \includegraphics[width=\textwidth]{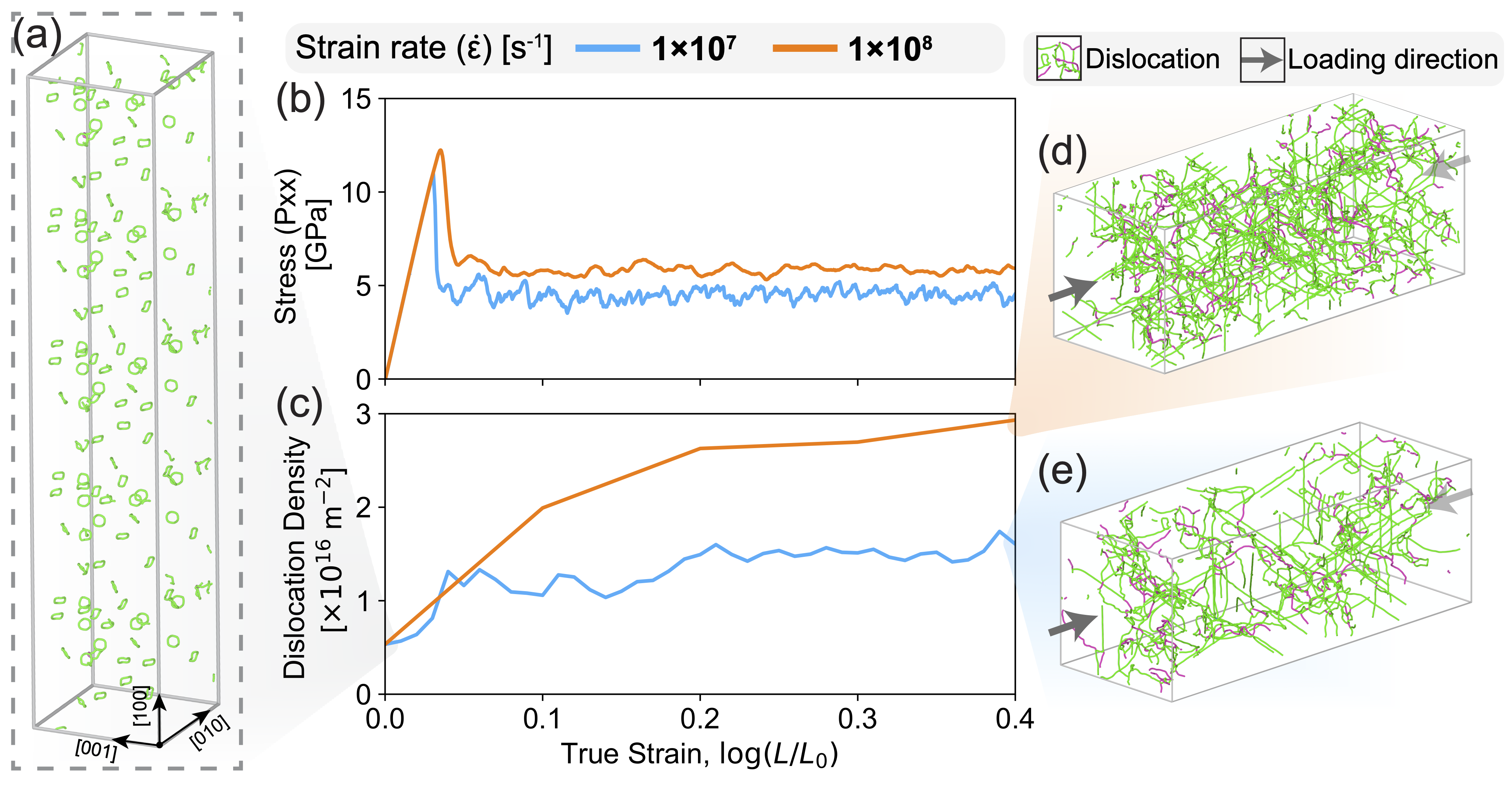}
\end{tabular}
\end{center}
  \caption{\label{fig:Fig1} (a) Single crystal of tungsten initiated with $1/2\langle111\rangle$ prismatic dislocation loops. (b,c) Stress and dislocation density versus true strain curves as a response of compression along [100] at different strain rates. The blue and orange curves represent strain rates of $10^7$ and $10^8$~s$^{-1}$, respectively. (d,e) Snapshots of dislocation networks developed post compression at a true strain of 0.4.}
\end{figure}

Pre-existing dislocations are introduced in our atomistic samples using the following procedure. First, the atomistic single-crystals are seeded with $1/2\langle111\rangle$ hexagonal-shaped prismatic dislocation loops (see Fig.~\ref{fig:Fig1}(a)) following the approach detailed in Ref.~\cite{zepeda2017probing}. After insertion of the loops, the crystals are annealed and equilibrated at the target deformation temperature under zero pressure using the {\it nph} barostat and the {\it langevin} thermostat.
The crystals are then pre-deformed under compression at various straining rates in the range $10^6-10^8$~s$^{-1}$ along the $X=[100]$ direction until a strain of 0.4. This enables us to obtain initial microstructures with fully developed dislocation networks of various dislocation densities, Figs.~\ref{fig:Fig1}(d,e), as the rate of dislocation multiplication from the initial loops directly depends on the straining rate \cite{zepeda2017probing,bertin2023crystal}.

Subsequent uniaxial tensile deformation was applied along the $Z=[001]$ direction of the simulation frame at various constant true straining rates. During deformation, the {\it nph} barostat and the {\it langevin} thermostat were used to maintain a uniaxial stress state and a constant target temperature.

\subsection{Experimental Methods}
99.95\% 25 $\mu$m thick tungsten foil was pre-machined into dog-bone shaped specimen with 3 mm by 0.1 mm gauge length and width, by Goodfellow Cambridge Ltd. These specimens were either fractured ex situ or prepared for \textit{In situ} testing using focused ion beam (FIB) milling (Quanta 3D, Thermo Fischer Scientific, USA). A region near the center of the specimen was thinned, using 30 keV Ga+ ions, to act as both a notch and thin transparent region for imaging during \textit{In situ} tensile testing of the entire sample. Tensile testing was performed using a custom uniaxial straining TEM holder where displacement is applied via a stepper motor with a crosshead speed set to 1 $\mu$m s$^{-1}$.  A JEOL 2010Plus LaB6 transmission electron microscope (TEM) was used for imaging. \textit{In situ} micro-tensile testing was performed using Hysitron PI 95 TEM Picoindenter (Bruker Corporation, USA). Micro-tensile specimens were also prepared using focused ion beam milling.  Micro-tensile testing was performed under displacement control loading at 2 nm s$^{-1}$.

\section{Acknowledgments}

{
This work was performed under the auspices of the U.S. Department of Energy (DOE) by the Lawrence Livermore National Laboratory (LLNL) under Contract No. DE-AC52-07NA27344. T.F. and O.H. acknowledge support from the U.S. DOE, Office of Science under an Office of Fusion Energy Sciences Early Career Award. F.A. would like to acknowledge support from the U.S. Army Research Office Award W911NF2310406. S.D. acknowledges support from the U.S. DOE, Office of Science under an Office of Fusion Energy Sciences E-SC0024690. Computing support for this work comes from the LLNL Institutional Computing Grand Challenge program.
An award of computer time was also provided by the INCITE program. This research used resources of both the Argonne and Oak Ridge Leadership Computing Facilities, which are DOE Office of Science User Facilities supported under contracts DE-AC02-06CH11357 and DE-AC05-00OR22725, respectively. The authors thank Pedro Borges and David Cook for helpful discussions.

}

\bibliographystyle{elsarticle-num} 
\bibliography{arxiv-refs}





\end{document}